\documentclass[graybox]{svmult}

\usepackage{mathptmx}       
\usepackage{helvet}         
\usepackage{courier}        
\usepackage{type1cm}        
\usepackage{makeidx}         
\usepackage{graphicx}        
\usepackage{multicol}        
\usepackage[bottom]{footmisc}

\usepackage{epstopdf}

\makeindex             

\begin{document}

\title*{Constraints on model atmospheres from
complex asteroseismology of the $\beta$ Cephei stars }
\author{Szewczuk Wojciech, Walczak Przemys{\l}aw and Daszy\'nska-Daszkiewicz Jadwiga}
\institute{Szewczuk W., Walczak P., Daszy\'nska-Daszkiewicz J. \at Instytut Astronomiczny, Uniwersytet Wroc{\l}awski,
Kopernika 11, 51-622 Wroc{\l}aw, Poland, \email{szewczuk@astro.uni.wroc.pl, walczak@astro.uni.wroc.pl, daszynska@astro.uni.wroc.pl}} 
%
%
\maketitle

\abstract*{Using the method termed complex asteroseismology, we derive constraints
on model atmospheres, in particular, on the NLTE effects.
We fit simultaneously pulsational frequencies
and the corresponding values of the nonadiabatic complex parameter $f$
for the four $\beta$ Cephei stars: $\theta$ Oph, $\nu$ Eri, $\gamma$ Peg and 12 Lac.
The LTE Kurucz models and the BSTAR2006 NLTE models are tested.}

\abstract{Using the method termed complex asteroseismology, we derive constraints
on model atmospheres, in particular, on the NLTE effects.
We fit simultaneously pulsational frequencies
and the corresponding values of the nonadiabatic complex parameter $f$
for the four $\beta$ Cephei stars: $\theta$ Oph, $\nu$ Eri, $\gamma$ Peg and 12 Lac.
The LTE Kurucz models and the BSTAR2006 NLTE models are tested.}

\section{Complex asteroseismology}
\label{sec:1}


We compute seismic models which fit centroid frequencies for different values
of mass, chemical composition and the core overshooting parameter.
From this set of models we choose those which reproduce the nonadiabatic parameter $f$
using the method of \cite{Daszynska}.
The $f$-parameter describes the ratio of the bolometric flux perturbation to the radial
displacement at the photosphere level
and its theoretical values are obtained from linear nonadiabatic theory of stellar pulsation.
All computations were obtained with the OPAL opacities. Two chemical mixtures were adopted:
A04 (\cite{Asplund04}) for $\nu$ Eri and $\theta$ Oph, and AGSS09 (\cite{Asplund09}) for 12 Lac and $\gamma$ Peg. 
The empirical values of $f$ were determined with the LTE Kurucz (\cite{Kurucz}) models and the NLTE model
atmospheres (\cite{Lanz}). Two values of the microturbulent velocity, $\xi_t$, were considered.

\section{Constraints on model atmospheres}
\label{sec:2}

The empirical values of the nonadiabatic  $f-$parameter of the $\beta$ Cep stars
 are sensitive to the model atmospheres (\cite{Daszynska}). In the case of 12 Lac  and $\gamma$ Peg, the values of $f$ obtained with the NLTE models
  reproduce the theoretical counterparts corresponding to seismic models with lower values of $\alpha_{ov}$.
  For 12 Lac, the minimum of $\chi^2$, as results from fitting the photometric amplitudes and phases, occurs for the NLTE atmospheres (see Table \ref{tab:1}).
  In the case of $\nu$ Eri and $\gamma$ Peg, the minimum of $\chi^2$ is reached  for the LTE models with
  the microturbulent velocity of $\xi_t$=8 km/s. On the other hand, for $\nu$ Eri the error box of the $f$-parameter
  for two dominant frequencies calculated with the NLTE models encloses more seismic models than with the LTE ones.
  We did not get any preference of model atmospheres for $\theta$ Oph, most probably because of the least accurate data.
  In Table\,1 we give the empirical values of $f$ at the minimum of $\chi^2$ for the dominant modes of the four studied stars. 
  These values were determined for various model atmospheres.

%
%
%
\begin{table}
\caption{The empirical values of $f$ for dominate modes at the minimum of $\chi^2$. For each star
calculations were performed for the central values of the observational error box.
Model of atmospheres are coded the same as in Kurucz \cite{Kurucz} and Lanz \& Hubeny \cite{Lanz}.}
\label{tab:1}       
%
%
\begin{tabular}{c | c| c| c| c| c|c | c| c| c| c| c} \hline
\hline\noalign{\smallskip}
freq        &     $f_{\rm R}$      & $f_{\rm I}$           & ~$\ell$~ &  $\chi^2$ & atm & freq      &     $f_{\rm R}$      & $f_{\rm I}$    & ~$\ell$~ &  $\chi^2$ & atm\\
 $[$c/d$]$       &        &            &  &   &  &   $[$c/d$]$    &         &    &  &   & \\
\hline
\multicolumn{6}{c}{ \textbf{12 Lac}}  &  \multicolumn{6}{|c}{ \textbf{$\gamma$ Peg}} \\
\hline
                     &           -8.66$\pm$0.38 &           -1.28$\pm$0.38&          &      12.9  & BGv2  &  &           -8.58$\pm$0.11 &            1.28$\pm$0.11&          &       5.84 & BGv2 \\
       5.1790 &           -9.17$\pm$0.45 &           -1.32$\pm$0.45&        1 &      16.1  & Kp00k2 &        6.5897 &           -8.89$\pm$0.09 &            1.31$\pm$0.09&        0 &       3.9  & Kp00k2 \\
                     &           -9.53$\pm$0.66 &           -1.38$\pm$0.66&          &      31.5  & Kp00k8  &                      &           -8.84$\pm$0.06 &            1.28$\pm$0.06&          &       1.5  & Kp00k8\\
\hline
\multicolumn{6}{c}{ \textbf{$\nu$ Eri}}   &  \multicolumn{6}{|c}{ \textbf{$\theta$ Oph}} \\
\hline
                     &           -8.88$\pm$0.39 &            0.75$\pm$0.40&          &      59.1  & BGv2 &                        &          -11.16$\pm$2.18 &            1.45$\pm$2.22&          &       3.5  & BGv2 \\
       5.7633 &           -9.24$\pm$0.37 &            0.76$\pm$0.38&        0 &      50.7  & Kp00k2 &          7.1160 &          -12.31$\pm$2.42 &            1.58$\pm$2.46&        2 &       3.6  & Kp00k2\\
                     &           -9.30$\pm$0.20 &            0.75$\pm$0.21&          &      14.4  & Kp00k8 &                        &          -12.90$\pm$2.61 &            1.67$\pm$2.65&          &       3.7  & Kp00k8\\
\hline
\end{tabular}

\end{table}

\section{Conclusions}
There is a prospect for deriving constraints on model atmospheres
from complex asteroseismology of the $\beta$ Cephei stars.
We got some for $\nu$ Eri, 12 Lac and $\gamma$ Peg. In the case of 
$\theta$ Oph, more accurate data are needed. Moreover, to achieve
a better consistency between theory and observations,
models of stellar atmospheres have to be still developed
and computed in more detail.

%

%

\end{document}